\documentstyle[prl,aps,multicol,epsfig,psfig]{revtex}

\begin{document}

\title{On the Cooling of  Electrons in a Silicon Inversion Layer}

\author{O.\ Prus, M.\ Reznikov, U.\ Sivan, }
\address{Dep. of Physics and Solid State Institute, Technion-IIT, Haifa 32000,
Israel}
\author{and V.\ Pudalov}
\address{P.\ N.\ Lebedev Physics Institute, 119991 Moscow, Russia}
\maketitle
\begin{abstract}

Cooling of two-dimensional electrons in silicon-metal-oxide
semiconductor field effect transistors is studied experimentally.
Cooling to the lattice is found to be more effective than expected
from the bulk electron-phonon coupling in silicon. The extracted
heat transfer rate to phonons at low temperatures depends
cubically on electron temperature, suggesting that another
coupling mechanism (such as piezoelectric coupling, absent in bulk
silicon) dominates over the deformation potential. According to
our findings, at 100\,mK, the electrons farther than $\sim
100~\mu$m from the contacts are mostly cooled by phonons. Using
long devices and low excitation voltage we measure electron
resistivity down to electron temperature $\sim 100$\,mK and find
that some of the ``metallic'' curves, reported earlier, turn
insulating below $\sim 300~m$K. This finding renders the
definition of the proposed 2D metal-insulator transition
questionable. Previous low temperature measurements in silicon
devices are analyzed and thumb rules for evaluating their electron
temperatures are provided.

\end{abstract}

\begin{multicols}{2}

Since the scaling theory for non-interacting electrons was
constructed in the late seventies\cite{gang4}, it was conjectured
that a realistic two-dimensional electron system is insulating in
the sense that its resistance diverges at low temperatures due to
quantum interference and interaction effects. As the temperature
is reduced, a logarithmic resistance increase due to weak
localization is expected, followed by an exponential resistance
divergence once strong localization commences\cite{gang4}. It was
later suggested\cite{Fin1} that strong Coulomb interaction may
counteract localization and lead to the existence of a two
dimensional metallic state and a metal-insulator transition  at
zero temperature.

Early experiments on various types of two dimensional electron
gases (2DEG) supported localization. For a comprehensive review of
early theoretical and experimental results see
Refs.~\cite{Rama,aa}. It was therefore unexpected when the
existence of a two dimensional metallic phase in high mobility
silicon Metal Oxide Semiconductor Field Effect Transistors
(MOSFET) was reported in 1994\cite{rhoT1,rhoT2}. The ultimate
signature of such a metallic phase is resistance saturation  to
some residual value as the temperature approaches zero
\cite{AbrKravSar,akk}. Since $T=0$ is experimentally inaccessible,
all claims for a metallic phase rely on extrapolation from finite
temperatures. Two traps lurk an experimentalist in this procedure:
(a) The assumption that the observed resistance saturation indeed
persists to zero temperature may turn wrong. (b) The electron
temperature may (and does indeed) depart from the lattice or bath
temperature, $T_b$, as $T_b\rightarrow 0$. The latter point is
particularly acute in silicon MOSFETs due to the high intrinsic
contact resistance and the weak electron-phonon coupling.

The experiments described here focus on the second point. Using
the sample itself as a thermometer we analyze electron cooling as
a function of mixing chamber temperature and excitation voltage.
The main findings are: (a) Electron cooling to the lattice is more
effective than expected from the known bulk electron-phonon
coupling in silicon. Power dissipation is given by $aT^3+bT^5$
with $a=2.2\times 10^{-8}~{\rm W/K^3 cm^2}$ and $b=5.1\times
10^{-8}~{\rm W/K^5 cm^2}$ and, hence, at temperatures below $\sim
0.6$~K, dominated by the $T^3$ term\cite{noteT3.3}. (b) Hot
electron diffusion (ED) to the contacts is unimportant in large
devices. In short Si MOSFETs, according to our estimates, ED
provides cooling or heating depending on the contact resistance.
(c) Comparing cooling by phonons with cooling by ED at e.g.,
100\,mK, we find the latter mechanism is typically important only
for electrons closer than $\sim 100\,\mu{\rm m}$ to the contacts.
In longer devices, cooling proceeds through phonons. (d) Cooling
the electrons down to ~100\,mK solely by phonons requires an
excitation power estimated to be less than $2\times 10^{-11}{\rm
W/cm^2}$. For a 1\,mm square of 2DEG with 30\,K$\Omega$
resistivity this figure implies an excitation voltage below
$80\,{\rm \mu V}$. A $50\,{\rm \mu m}$ square with the same
resistance requires an excitation voltage below $4\,{\rm \mu V}$.

Four terminal, lock-in ac measurements at 11 Hz were carried out
on samples similar to those used in Refs.\cite{rhoT1,rhoT2}. The
samples were 5\,mm long, 0.8\,mm wide Hall bars with 2.5\,mm
separation between potential probes. The oxide thickness was
200\,nm. Battery powered electrometer preamplifiers were employed
to minimize electromagnetic interference and spurious offset bias
across the potential probes. With 300pF cable capacitance, samples
up to a few ${\rm M\Omega}$ total resistance were measured with
negligible ``out of phase'' signal. The excitation voltage across
the whole sample was maintained below $100\, {\rm \mu V}$ to
guarantee tolerable heating (this figure is justified below). In
order to reduce the contact resistance we applied a larger
positive front gate voltage and $-15\,V$ substrate bias, thus
maintaining the same carrier concentration range. The substrate
bias depletes both the channel and the contacts, but it's effect
on the latter is smaller due to heavy doping in the contact
regions. This way we were able to reduce the contact resistance
below $20\,{\rm K\Omega}$ for all ``metallic'' densities. The
substrate bias results in a moderate reduction of the peak
mobility from e.g., ${\rm 3\times 10^4\,cm^2/Vs}$ at zero
substrate bias to ${\rm 2.1\times 10^4cm^2/Vs}$ (both measured at
$T=0.3$\,K ). The mobility reduction is attributed to enhanced
surface scattering due to shift of the 2DEG towards the ${\rm
Si-SiO_2}$ interface. The substrate bias was changed in the
presence of red LED illumination, which heats the crystal, thus
facilitating acceptor recharging in the depletion layer. The high
contact resistance problem is further complicated below 1.2K by
superconductivity of the Al contact metallization. All our ``zero
magnetic field'' measurements were, therefore, carried out in the
presence of a 0.01\,T parallel magnetic field which quenched the
Al superconductivity.

\begin{figure}
\centerline{\psfig{figure=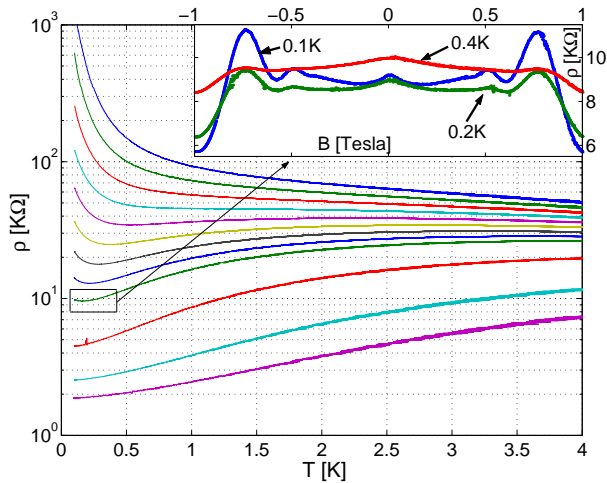,width=230pt} }
\begin{minipage}{3.2in}
 \caption{Temperature dependence of the resistivity. Densities from top to bottom
  are $n=1.20$ to 1.44 (in 0.03 steps), 1.56, 1.8  and  2.04${\rm\times 10^{11}~cm^2/Vs}$.
  Inset: Magnetoresistance at   $n={\rm 1.44\times10^{11}~cm^2/V s}$. The weak localization
  negative magnetoresistance peak and the onset of Shubnikov - de Haas oscillations are clearly seen.
  The peak width decreases and the oscillations amplitude increases
  without saturation as $T_b$ decreases to 0.1K.}
 \label{SiR(T)h}
  \end{minipage}
\end{figure}

The resistivity  vs bath temperature for a $ 2.1\times10^4~{\rm
cm^2/V s}$ peak mobility  sample at various densities is depicted
in Fig.~\ref{SiR(T)h}. For temperatures above 0.4\,K the traces
are similar to those reported in Refs.~\cite{rhoT1,rhoT2}. At low
densities the sample is insulating, while at higher densities the
resistance decreases as the temperature is reduced, a behavior
previously assigned to a two dimensional metal. For $n=1.29\times
10^{11} {\rm cm^{-2}}$ the resistance is practically independent
of $T$ in a wide temperature range, $0.4\div4$\,K. Such a
phenomenon was previously identified as a metal to insulator
transition (MIT) \cite{rhoT1,rhoT2,AbrKravSar}. Upon cooling the
sample below 0.4K, we find a novel result; the resistance of some
of the ``metallic'' curves turns upward, showing an insulating
behavior in all three devices studied by us.

\begin{figure}
\centerline{\psfig{figure=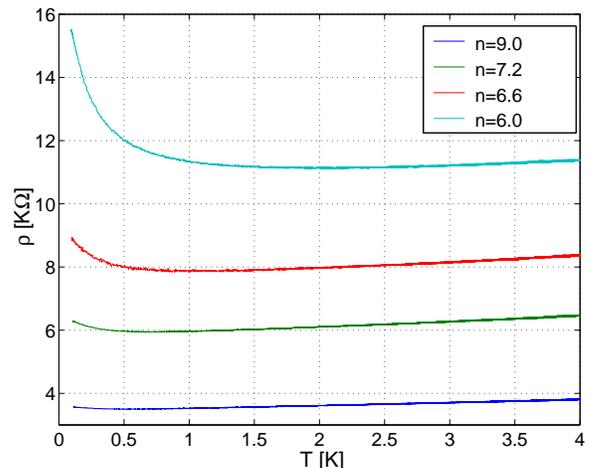,width=220pt}}
\begin{minipage}{3.2in}
 \caption{Resistivity dependence upon temperature for a
 $2.5 \times 10^3~{\rm cm^2/Vs}$ peak mobility sample. Densities given in
 $10^{11}~{\rm  cm^2/Vs}$ units.}
 \label{SiR(T)l}
 \end{minipage}
\end{figure}

Measurements on a lower mobility sample with $2.5 \times 10^3 {\rm
cm^2/Vs}$ peak mobility (Fig.~\ref{SiR(T)l}) at 0.3 K yield
results similar to those reported earlier\cite{akk,mauterndorf}
for the same type of samples. As the temperature decreases,  the
low mobility sample displays a pronounced insulating behavior at
all measured densities, above $6\times 10^{11}{\rm cm^{-2}}$. A
relatively small resistivity drop with cooling is again followed
by a resistivity increase at lower temperatures. Such curves have
been observed earlier for 2D holes in Si-Ge \cite{senz}, 2DEG on
vicinal Si-surfaces \cite{safonov}, high mobility Si MOSFETs at
higher densities, $n> 2\times 10^{12}{\rm cm^{-2}}$
~\cite{gmax,app}, and in 2D systems on other
materials~\cite{Simons,ribeiro}. We attribute this resistance
upturn to interference and Coulomb interaction effects that grow
as the temperature is reduced. The interference is clearly
manifested in the weak localization magnetoresistance curves
depicted in the inset to Fig.~1.

To gain insight into the electron cooling mechanisms we studied
the resistivity dependence upon excitation. We applied a small ac
signal superimposed on a dc bias and measured the differential
resistivity as a function of this bias. The ac excitation was
typically about 100\,pA, generating less than ${\rm 65~\mu V}$ ac
voltage drop across the 5\,mm sample. The analysis assumes the
resistivity depends on the electronic temperature alone and uses
the sample resistivity as a thermometer. We chose to use the low
mobility sample due to its moderate resistance and monotonic
temperature dependence, which facilitate accurate ac measurements.

\begin{figure}
\centerline{\psfig{figure=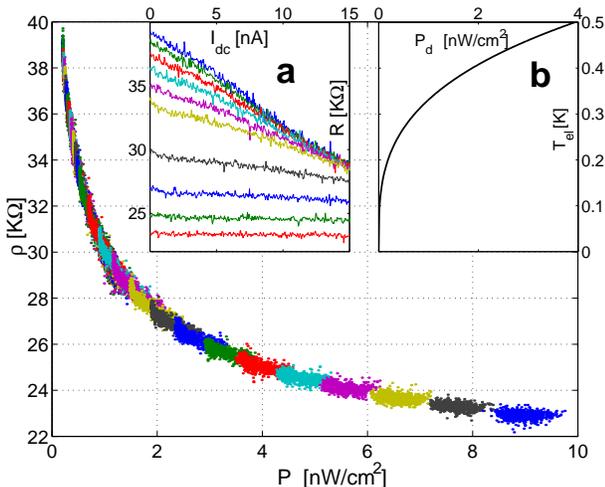,width=230pt}}
\begin{minipage}{3.2in}
 \caption{Channel resistance scaling for the low mobility sample. The
 abscissa depicts $P=P_{d}+aT^3+bT^5$.
  The fitting parameters are: $a=2.2 \times 10^{-8}~{\rm W/K^3 cm^2}$ and
  $b=5.1 \times 10^{-8}~{\rm W/K^5  cm^2}$. Colors correspond to the same curves as in inset a.
  Insets: a) resistivity versus dc current at different base temperatures: 90,
  120, 150, 180, 210, 240, 330, 420, 510, 600~mK.
  b) the electronic temperature calculated using eq.~\protect\ref{heatmodel}
   as a function of dissipated power for $T_b=0,~P_0=0$.}
\label{Rel(P)}
\end{minipage}
\end{figure}

Inset (a) to Fig.~\ref{Rel(P)} depicts the channel resistivity
versus dc current at various bath temperatures. The overall
dissipated power per unit area of the 2DEG consists of Joule
heating, $P_{d}=J^{2}\times \rho$, by the driven current and
heating by spurious excitations, $P_{0}$, due to e.g., absorption
of electromagnetic radiation. The heat is dumped to phonons and to
the contacts via electron diffusion. We ignore for the moment
electron diffusion - an assumption to be later justified for our
samples. In the general case, electron-phonon interaction may be
due to both piezoelectric and deformation potential coupling with
their characteristic $T^3$ and $T^5$ dependencies\cite{Karpus},
respectively. Similar functional form was recently found for holes
in SiGe heterostructures~\cite{SiGe}. We therefore characterize
the electron phonon interaction by the coupling parameters $a$ and
$b$.

\begin{equation}\label{heatmodel}
  P_{d}+P_{0}=a(T_{el}^3-T_b^3)+b(T_{el}^5-T_b^5)
\end{equation}

We solve Eq.~1 to find the electron temperature, $T_{el}$, and fit
$a$ and $b$ in such a way that all $R(T_{el})$ data in inset (a)
to Fig.~3 collapse to a single curve. Figure~\ref{Rel(P)} shows
the best fit obtained with $a=2.2 \times 10^{-8} {\rm W/K^3 cm^2}$
and $b=5.1 \times 10^{-8} {\rm W/K^5 cm^2}$. $P_0$ cannot be
determined in this scaling procedure. The piezoelectric
contribution is, hence, found to dominate electron - phonon
coupling at all temperatures below 0.6\,K. This effect is
unexpected since a Si lattice possesses inversion symmetry and,
hence, should display deformation potential coupling only. We
believe the piezoelectric contribution, found in our experiment,
originates from the ${\rm Si-SiO_2}$ interface which lacks
inversion symmetry.

Heat transfer due to ED can be incorporated into the model using
the Wiedemann-Franz law \cite{note_WF}. The power balance reads

\begin{eqnarray}\label{model}
 a(T_{el}^3-T_b^3)+b(T_{el}^5-T_b^5)=\nonumber\\
 P_d +P_0+ \frac{1}{6}\left(\frac{\pi k_{B}}{e}\right)^{2}
  \sigma\ \frac{\partial^2{T^2}}{\partial x^2},
\end{eqnarray}

where $\sigma$ is the sample conductivity. At temperatures
$T_{el}\ll 0.6{\rm K}$ the deformation potential contribution can
be ignored. Defining $T_0^3=T_b^3+(P_d+P_0)/a$ and

\begin{equation}\label{lambda}
\lambda=\frac{\pi k_B}{3e}\sqrt{\sigma /aT_{0}},
\end{equation}
Eq.~\ref{model} takes the dimensionless form

\begin{equation}\label{unitless}
\Theta^3 -1=\frac {3} {2} \frac{\partial^2 \Theta^2}{\partial
\chi^2}
\end{equation}

where $\Theta\equiv T_{el}/T_0$ and $\chi\equiv x/\lambda$.
$\lambda$ sets the characteristic length scale on which the
$T_{el}$ approaches $T_0$. The only temperature scale in the
problem is $T_0$, the electron temperature far away from the
contacts, where cooling is dominated by phonons. Equation
\ref{unitless} can be integrated exactly. Away from the contacts,
where $\Theta\approx 1$, $\Theta\approx 1+exp(-\chi)$. For
$\Theta\gg1,~\Theta(\chi)\approx\Theta_c/(1+\sqrt{(\Theta_c/30})\chi)^2$,
where $\Theta_c$ is the normalized electron temperature near the
contact.

For $T_0=100\,{\rm mK}$ and $\sigma=3.3\times 10^{-5}{\rm
Ohm^{-1}}$ (typical critical region conductance), $\lambda\approx
110\mu{\rm m}$. For our 5mm long samples, ED is, hence, negligible
and Eq.~\ref{heatmodel} can be safely used to estimate $T_0$. The
resulting electron temperature as a function of dissipated power,
is plotted in inset (b) to Fig.~\ref{Rel(P)}. Assuming that the
electron-phonon coupling does not depend significantly on impurity
scattering we use the same parameters to analyze the high mobility
sample data. Equation \ref{heatmodel} shows that the overheating
due to the measurement current is negligible and does not exceed 1
mK for a 50 mK base temperature. We have used the
magnetoresistance curves shown in the inset to Fig.~1 to estimate
the lowest sample temperature by fitting the curves to weak
localization theory. The extracted dephasing rate scales linearly
with temperature, in agreement with theory\cite{aa}, down to the
lowest analyzed temperature, $T_b=100$\,mK. Although we use weak
localization theory for resistivities beyond its applicability,
the linear dependence of the dephasing rate upon bath temperature
indicates that electrons cool down to $T_{el}\sim 100$\,mK.

We turn now to investigate the implications of our results to
previous measurements. Surprisingly, there are only a few
published experiments below 300\,mK. Those experiments were
carried out on two types of samples. The first type, in which the
MIT was originally observed \cite{rhoT1,rhoT2}, is a conventional
MOSFET that, in the lack of back gating, suffers from high contact
resistance, especially at low electron densities and temperatures.
These samples are several mm long, and heat transfer by electron
diffusion to the contacts can be safely neglected. In the
interesting regime, where the resistance drop is significant, the
contact resistance is in the ${\rm M\Omega}$ range. Such a high
contact resistance dictates either dc or very low frequency
measurements which require, due to $1/f$ noise in the amplifiers
and the sample, substantial excitation currents. Overheating is
evident in both $R$ vs $T$ and $\ln R$ vs $1/T$ dependencies (see
e.g., Figs.~1 and ~5 in ~\cite{rhoT2} and Fig.~3 of~\cite{prb92}),
even in the insulating regime where it should be less significant.
In previous experiments\cite{weakloc,gm}, where $T_{el}$ was
measured in the ``metallic'' regime, it was never lower than
$\approx 200 \div 300$~mK.

\begin{figure}
\centerline{\psfig{figure=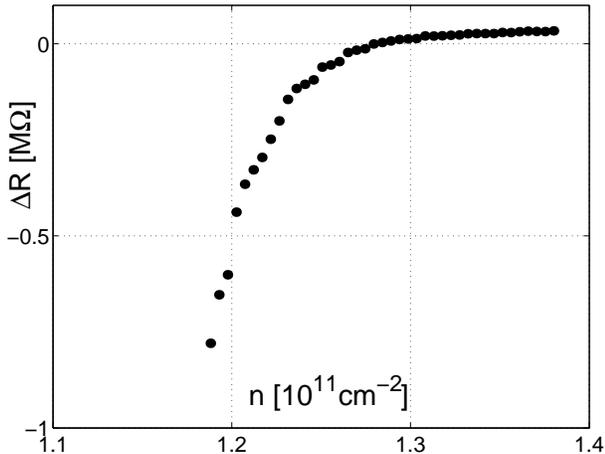,width=230pt}}
\begin{minipage}{3.2in}
\caption{The difference between the two terminal resistance,
$R_2$, and the ideal resistance, $\tilde{R_2}=\rho\times{L/W}$,
corresponding to a uniform sample temperature and zero contact
resistance. $T_b=80$~mK and the excitation voltage $\leq 65~\mu$V.
Note the difference changes sign at approximately the MIT
density.}
 \label{Rc(P)}
\end{minipage}
\end{figure}

In  another type of  samples\cite{heemskerk,KK}, the gates,
adjacent to the contacts, are separated from the gate over the
central part of the sample by $\approx 50$\,nm gaps. This
arrangement facilitates a higher carrier concentration in the
contact regions and, hence, a reduced contact resistance. In
contrast to the millimeters long samples, employed here and in
Refs.~\cite{rhoT1,rhoT2}, the effective sample length between the
gaps in Ref.~\cite{KK} is only $120 \mu m$. Substituting
$\lambda=60~{\rm \mu m}$ into Eq.~\ref{lambda} and solving for
$T_{0}$ we find that cooling is mostly provided by phonons for all
temperatures above $\sim 340$~mK. Since cooling by phonons is
suppressed only slightly faster than by ED ($T^3$ compared with
$T^2$), phonons, neglected in the analysis of Ref.~\cite{KK} done
in Ref.~\cite{akk}, are important also at lower temperatures.
Electron diffusion from the contacts may either cool or warm
electrons in the central part of the sample, depending on the
voltage drop, $U_c$, across the $\approx 50$\,nm gap. For a given
electron temperature near the contact, $T_c$, one can estimate the
voltage drop, $U_c$, across the gap resistance, $\rho_c$,  for
which the net heat flow between the contact and the sample
vanishes. Using Wiedemann-Franz law, the heat flow balance reads

\begin{equation} \label{c1}
U_c^2/2\rho_c=\frac {\pi^2} {6\rho_c} \left(\frac{k_b}
{e}\right)^2(T_c^2-T_b^2),
\end{equation}

For $T_b=0$ one obtains $U_c/T_c=\left(\pi k_b/\sqrt 3 e\right)
\sim156 \mu {\rm V/K}$, independent of $\rho_c$. For lower
(higher) voltages across $\rho_c$, electrons are cooled (warmed)
by the contacts. For 100~mK, the bias at which the contacts
neither heat nor cool the sample equals 15.6~$\mu {\rm V}$. At
that bias sample cooling is provided by phonons. Using Eq.~1 we
find that 100~mK electron temperature in a 30${\rm k\Omega}$ per
square sample requires less than $\sim 4~\mu V$ voltage drop per $
50~\mu$m square, which for the sample in Ref.~\cite{KK},
translates into $10~\mu V$ between the gate gaps. A 100~mK
electron temperature throughout the sample thus requires less than
$\sim 40~\mu$V (50~mK requires  $< 20 \mu$V ) bias across the
whole sample (neglecting the voltage drop across the high density
contacts). Electron overheating in the contact regions occurs in
our samples, as illustrated in Fig.~\ref{Rc(P)}. The contact
regions are always more resistive than the rest of the sample in
the ``metallic'' regime, $n>{\rm 1.29 \times 10^{11} cm^{-2}}$,
and less resistive in the insulating one, $n<{\rm 1.29 \times
10^{11} cm^{-2}}$, indicating they are warmer than the rest of the
sample.

In summary, we found that the cooling of electrons by phonons in
Si MOSFETs is substantially more effective than that expected from
the bulk electron-phonon coupling. Long samples are cooled by
phonons. Short samples are susceptible to heat transfer by ED from
the contacts. Low temperature measurements in short samples
require low excitations. Using 5 mm long devices, where electron
heating in the contacts can be neglected, we were able to measure
electron resistivity down to $\sim 100$\,mK temperatures. We found
that $\rho(T)$-curves in the well pronounced ``metallic'' regime,
corresponding to $(1.3-1.44)\times 10^{11}$\,cm$^{-2}$ densities
and a three fold resistance reduction with cooling, turn
insulating below a certain temperature. That temperature is $\sim
300$~mK at $\rho \sim h/e^2$ and decreases as the density is
increased and $\rho$ decreases. Our finding renders the definition
the 2D metallic phase and the metal-insulator phase transition
questionable. Further work is needed to determine the physics
behind the observed resistance upturn.

We have benefited greatly from valuable discussions with Y.~Yaish,
M.~Gershenson, A.~Finkelstein, A.~Stern, and T.\ M.\ Klapwijk.
This work was supported by the Israeli National Science
foundation, the DIP foundation, INTAS, NSF, NATO Scientific
Program, RFBR, and the Russian programs: ``Physics of
Nanostructures'', ``Statistical Physics'', ``Integration'', and
``The State Support of Leading Scientific Schools''.

\end{multicols}

\begin{references}
\bibitem{noteT3.3}This form is equivalent to the empirical law,
$\sim T^{3.3}$ over the range $T=0.2\div 4$\,K.
\end{references}
\end{document}